\pdfoutput=1
\documentclass[conference,a4paper,10pt]{IEEEtran}

\usepackage[latin1]{inputenc}
\usepackage{times,amsmath}
\usepackage{pstool}
\usepackage{subfigure}
\usepackage{multirow}
\usepackage{enumerate}
\usepackage{graphicx}
\usepackage{MnSymbol}
\usepackage{stfloats} 
\usepackage[table]{xcolor}
\usepackage[square, comma, sort&compress, numbers]{natbib}
\usepackage{hyperref}
\usepackage{algorithm,algorithmic}

\graphicspath{ {Figures/} }

\begin{document}

\title{Shared Secret Key Generation via Carrier Frequency Offsets}

\author{
\IEEEauthorblockN{
Waqas Aman\IEEEauthorrefmark{1},
Aneeqa Ijaz\IEEEauthorrefmark{1}, 
M. Mahboob Ur Rahman\IEEEauthorrefmark{1},
Dushanta Nalin K. Jayakody\IEEEauthorrefmark{2}\IEEEauthorrefmark{3},
Haris Pervaiz\IEEEauthorrefmark{4}
} 
\IEEEauthorblockA{
\IEEEauthorrefmark{1}Electrical Engineering department, Information Technology University (ITU), Lahore, Pakistan \\\ \{waqas.aman,aneeqa.ijaz,mahboob.rahman\}@itu.edu.pk }
\IEEEauthorblockA{
\IEEEauthorrefmark{2}School of Computer Science \& Robotics, National Research Tomsk Polytechnic University, Russia (nalin@tpu.ru) }
\IEEEauthorblockA{
\IEEEauthorrefmark{3}School of Postgraduate Studies and Research,  Sri Lanka Technological Campus, Sri Lanka }
\IEEEauthorblockA{
\IEEEauthorrefmark{4}School of Computing and Communications, Lancaster University, UK (h.b.pervaiz@lancaster.ac.uk) }
}

\maketitle

\begin{abstract} 

This work presents a novel method to generate secret keys shared between a legitimate node pair (Alice and Bob) to safeguard the communication between them from an unauthorized node (Eve). To this end, we exploit the {\it reciprocal carrier frequency offset} (CFO) between the legitimate node pair to extract common randomness out of it to generate shared secret keys. The proposed key generation algorithm involves standard steps: the legitimate nodes exchange binary phase-shift keying (BPSK) signals to perform blind CFO estimation on the received signals, and do equi-probable quantization of the noisy CFO estimates followed by information reconciliation--to distil a shared secret key. Furthermore, guided by the Allan deviation curve, we distinguish between the two frequency-stability regimes---when the randomly time-varying CFO process i) has memory, ii) is memoryless; thereafter, we compute the key generation rate for both regimes. Simulation results show that the key disagreement rate decreases exponentially with increase in the signal to noise ratio of the link between Alice and Bob. Additionally, the decipher probability of Eve decreases as soon as either of the two links observed by the Eve becomes more degraded compared to the link between Alice and Bob.

\end{abstract}

\section{Introduction}
\label{sec:intro}

Physical-layer security has its roots in 1950's when Shannon argued that perfect secrecy is possible provided that the entropy of the secret key is greater than the entropy of the to-be transmitted message \cite{shannon:BSTJ:1949}. Later on, Wyner, in his influential work introduced the notion of Gaussian wiretap channel to compute the so-called secrecy capacity in additive white Gaussian noise (AWGN) channels \cite{Wyner:BLTJ:1975}. Csiszar \cite{Csiszar:TIT:1978} then extended the notion of secrecy capacity to the wireless fading channels. Maurer \cite{Maurer:TIT:1993} was first to suggest to extract shared secret keys from a common source of randomness. Nevertheless, until last decade, the world had been accustomed to using higher-layer cryptographic protocols for authentication/security purposes. More recently, there is a growing interest in designing algorithms at the physical layer so as to complement/improve the existing security mechanisms, see, e.g., \cite{Amitav:corr:2010},\cite{Shiu:WC:2011} for a quick overview of recent development in the field.

In the literature on physical layer security, two popular models exist: i) Wyner's wiretap model, and ii) Basic source model. Wyner's wiretap model assumes that eavesdropper is using a degraded version of the main channel, and utilizes channel coding to approach the secrecy capacity. Having said this, much work has been done to design channel coding schemes which meet the absolute limits of secrecy capacity \cite{Bloch:TIT:2008}. On the other hand, under the Basic source model, two legitimate nodes obtain multiple correlated observations from a shared random source. Both nodes then quantize their observations, do the information reconciliation \cite{Bloch:ITW:2006} (to eradicate the bit mismatch at both ends) followed by privacy amplification \cite{Bennett:TIT:1995} (to hash out the bits revealed during information reconciliation phase) to distil a shared secret key. 

For the Basic source model, researchers have exploited the random and reciprocal nature of wireless medium in single-antenna and multiple-antenna settings to generate shared secret keys \cite{Ye:TIFS:2010}, \cite{Ye:ISIT:2006}. Additionally, the feasibility of using the relays/friendly jammers to design high performance key generation algorithms is reported in \cite{Dong:TSP:2010}, \cite{Golla:infocom:2011}.

Apart from the medium, physical characteristics of the underlying device hardware can also be used for security, e.g., integrated circuits \cite{Lim:VLSI:2005}, oscillators \cite{Wang:ICC:2012,Mahboob:Globecom:2014,Mahboob:ICUWB:2015,Satyam:Arxiv:2017}, antennas \cite{Imai:JWIS:2006}, non-reciprocal hardware \cite{Mahboob:VTC:2017S} etc. In this paper, we exploit \textit{reciprocal carrier frequency offset} (CFO) between a node pair to generate secret keys shared between that node pair. To the best of authors' knowledge, there has been no work on this expect \citep{Wang:ICC:2012,Mahboob:Globecom:2014,Mahboob:ICUWB:2015}, which all use the CFO for authentication. 

The main contributions of this paper are two-fold: i) a novel algorithm which constructs shared secret keys from the noisy CFO estimates ii) key generation rate of the CFO process. 

The rest of this paper is organized as follows. Section II introduces the system model and the CFO models. The proposed, CFO based method for secret key generation is presented in Section III. Section IV studies the key generation rate of the CFO process. Section V provides some simulation results. Finally, Section VI concludes.

\section{System Model \& CFO Background}
\label{sec:sm}

\subsection{System Model}
\begin{figure}[ht]
\begin{center}
	\includegraphics[width=3in]{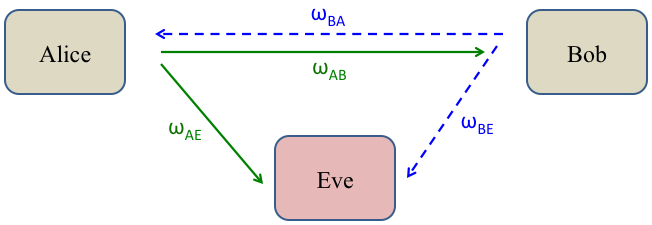} 
\caption{System model.}
\label{fig:sysmodel}
\end{center}
\end{figure}

The system model consists of three nodes, Alice, Bob and Eve. As shown in Fig. \ref{fig:sysmodel}, Alice and Bob make a legitimate node pair who intend to establish a secure wireless communication link in order to exchange confidential messages. Eve is a malicious node who passively eavesdrops in order to decipher the shared secret key being used by Alice and Bob. The legitimate node pair operates in half-duplex/time-division duplex (TDD) mode with $T$ seconds long time-slots. Specifically, in order to measure the CFO to learn a shared secret key, Alice and Bob exchange binary phase shift keying (BPSK)-modulated packets to each other. Finally, the center frequency of the channel is $\omega_c$ rad/sec.

\subsection{CFO is Reciprocal}
\begin{figure}[ht]
\begin{center}
	\includegraphics[width=3.5in]{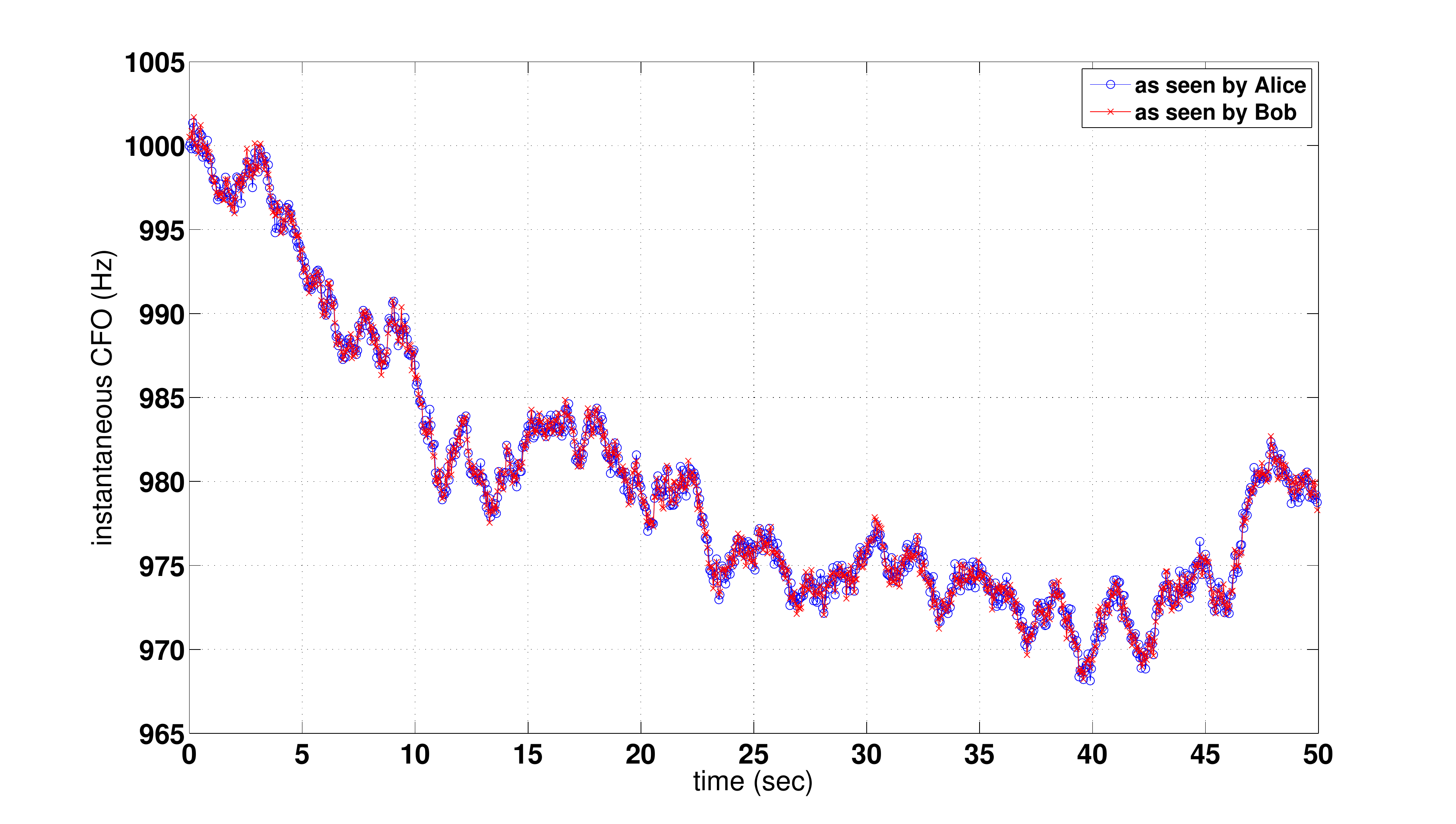} 
\caption{Experimental validation: CFO between a node pair is reciprocal.}
\label{fig:lo-offset}
\end{center}
\end{figure}

CFO is a measure of the speed of oscillations of a device's oscillator relative to that of another device. CFO arises due to manufacturing tolerance of oscillators; and may drift over time due to environmental/operating conditions. CFO is {\it reciprocal}: let $\omega_{AB} = \omega_A - \omega_B$ (rad/sec) be the CFO between "Alice and Bob", then reciprocity implies that $\omega_{AB}=-\omega_{BA}$. Therefore, the mutual CFO $\omega_{AB}$ can indeed be exploited by the legitimate node pair (Alice and Bob) to generate shared secret keys every once in a while. Fig. \ref{fig:lo-offset} plots the two CFO's ($f_{AB}=\omega_{AB}/2\pi$\footnote{We use the notation $\omega$ (rad/sec) and $f$ (Hz) in interchangeable manner throughout the rest of the paper.} and $-f_{BA}=-\omega_{BA}/2\pi$) against time. To obtain Fig. \ref{fig:lo-offset}, an experiment was set up whereby two GNU Radio/USRP based software-defined radios (SDR) exchanged unmodulated tones/sinusoids with each other in frequency-division duplex (FDD) fashion to measure the (time-varying) CFO in both directions. Fig. \ref{fig:lo-offset} verifies that the CFO is indeed reciprocal.

\subsection{CFO Models}
We now introduce the three fundamental models which govern the random, time-varying nature of the CFO.

{\bf Model M1: CFO is time-invariant.} Under this model, the CFO $\omega_{AB}$ is treated as a random variable with distribution $U(-2\pi \Delta,2\pi \Delta)$ where $\Delta$ could be derived from the parts-per-million (ppm) specs of the oscillators under consideration. This work considers the homogeneous case, i.e., when all the three nodes (Alice, Bob, Eve) of the considered system model use oscillators with same stability (ppm) specification. Let each of the three oscillators have an accuracy of $x$ ppm, then $\Delta=f_c \times x$ Hz ($f_c$ is the center frequency in MHz). 

The construction of the remainig two models, model M2 and model M3 is based upon the so-called Allan Deviation\footnote{Allan deviation is a well-known measure of frequency-stability of the oscillators \cite{Zucca:ITUFFC:2005}, \cite{Galleani:Metrologia:2008}.}. So, Allan deviation first. 

\begin{figure}[ht]
\begin{center}
	\includegraphics[width=3in]{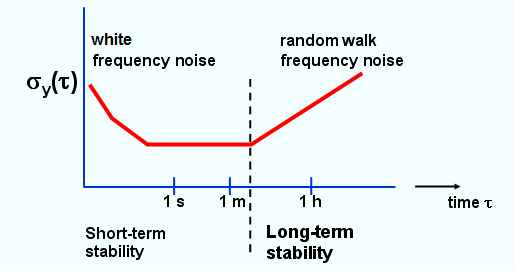} 
\caption{Allan deviation $\sigma_y(\tau)$ as a function of the observation interval $\tau$.}
\label{fig:avar}
\end{center}
\end{figure}

{\bf Allan deviation.} Fig. \ref{fig:avar} shows a typical plot of the Allan deviation $\sigma_y(\tau)$ against the observation interval $\tau$. Fig. \ref{fig:avar} indicates that there are two frequency-stability regions for the oscillators. In the short-term stability region (which lasts from few seconds to few minutes), white frequency noise dominates, while in the long-term stability region, random walk frequency noise dominates. 

{\bf Model M2: CFO is time-varying and memory-full.} Model M2 represents the long-term stability region of the Allan deviation curve. Here, aging/temperature effects cause the CFO to undertake a random walk over time \cite{Zucca:ITUFFC:2005}, \cite{Galleani:Metrologia:2008}:
\begin{equation}
	\label{eq:wAB}
	 \text{Model M2a:} \; \omega_{AB}(k+1) = \omega_{AB}(k) + n_{AB}(k)
\end{equation}
where $n_{AB}(k) \sim N(0,\sigma^2)$ is the random walk frequency noise. Let $t_k - t_{k-1}=T$ be the duration of a time-slot. Then, $\sigma^2=\omega_c^2 q_2^2 T$ where $q_2^2 = 5.51 \times 10^{-18}$ for USRP N200 radios \cite{Quitin:TWC:2013}. Re-arranging Eq. (\ref{eq:wAB}), we have: 
\begin{equation}
	\label{eq:inc}
	\text{Model M2b:} \; n_{AB}(k)=\omega_{AB}(k+1)-\omega_{AB}(k) 
\end{equation}
Note that the original stochastic process $\{\omega_{AB}\}_k$ of Eq. (\ref{eq:wAB}), Model M2a, is non-stationary auto-regressive moving average (ARMA) process, while the stochastic process $\{n_{AB}\}_k$ of Eq. (\ref{eq:inc}), Model M2b, is stationary with independent and identically distributed (i.i.d) elements.  

{\bf Model M3: CFO is time-varying and memoryless.} Model M3 represents the short-term stability region of the Allan deviation curve. Here, the CFO $\{\omega_{AB}\}_k$ is a memoryless random process. That is, the CFO stays constant for the slot duration $T$; moreover, the CFO realizations across the slots are i.i.d $U(-2\pi \Delta,2\pi \Delta)$. This model resembles closely the well-acclaimed block-fading model for wireless channels. 

At this point, some comments about the three CFO models are in order. Model M1, model M3 represent extreme/limiting cases whereby the CFO does not change at all, change independently during every time-slot, respectively. Furthermore, Model M1 represents an ideal oscillator (closest to which are the atomic clocks). On the other hand, all the commodity oscillators follow the Allan deviation curve which implies that they follow either model M2, or, model M3, depending upon the total time of their operation. Finally, we note that model M1 provides only 1 secret key during the life-time of an (ideal) oscillator; therefore, the rest of this paper will focus on model M2 and model M3 (and thus, commodity oscillators) only.

\section{The Proposed Method}
\label{sec:method}

Due to two-way communication between Alice and Bob, four CFOs are of interest: $\omega_{AB}, \omega_{BA}, \omega_{AE}, \omega_{BE}$  (see Fig. \ref{fig:sysmodel}). Appendix \ref{app:blind_est} describes a blind method for CFO estimation from BPSK-modulated data. Having obtained the noisy CFO estimates $\hat{\omega}_{AB}(k)=\omega_{AB}(k)+\nu_{AB}(k)$ and $\hat{\omega}_{BA}(k)=\omega_{BA}(k)+\nu_{BA}(k)$, Alice and Bob utilize them to generate secret keys. $\nu_{AB}(k)\sim N(0,\sigma_{AB}^2)$ ($\nu_{BA}(k)\sim N(0,\sigma_{BA}^2)$) is the estimation error at Alice (Bob). Specifically, with the CFO measurements in hand, the legitimate nodes need to do information reconciliation followed by privacy amplification. For information reconciliation, both nodes utilize linear block codes to exchange syndrome to eradicate the bit mismatch\footnote{Note that when the public discussion for information reconciliation is not feasible, each of the legitimate nodes (Alice and Bob) could do majority decision decoding at its end for authentication. That is, Bob authenticates Alice if the received secret key and the local key have at most $p$ ($0<p<n$) mismatches, where $n$ is the length of the shared secret key.}. For privacy amplification, universal hash functions could be used to hash out the information revealed. 

The essential steps of the proposed method are formally summarized below. Alice and Bob: 
\begin{enumerate}
\item exchange BPSK signals to perform blind CFO estimation on the received signals to get $\hat{\omega}_{BA}$, $\hat{\omega}_{AB}$, respectively. 
\item quantize their individual CFO estimates using equi-probable/uniform quantization to get $K_A$ and $K_B$, respectively. $K_A$ ($K_B$) is length-$n$ binary key at Alice (Bob).
\item do information reconciliation using linear block codes to construct reconciled keys $\mathcal{K}_A$ and $\mathcal{K}_B$, respectively. 
\end{enumerate}

Note that uniform quantization in step 2 results in some entropy loss, while information reconciliation in step 3 reveals some information as well (due to public discussion). Thus, both steps 2,3 reduce the secret bit rate (SBR) to some extent. 

{\bf Remark 1.} Since model M2 is an ARMA process, both Alice and Bob implement a linear Kalman filter (LKF) (after step 1 and before step 2) to effectively track the drifting CFOs. Each of the two LKFs is fed by the noisy CFO estimate (outputted by the blind estimation method) and yields the filtered CFO estimate. Note that the LKF is the best linear unbiased estimator of the CFO. Therefore, once LKF is converged, each legitimate node utilizes its filtered estimate to implement step 2 and step 3. More details on using the LKF to track the drifting CFOs could be found in \cite{Mahboob:Globecom:2014},\cite{Quitin:TWC:2013}.

\section{Key Generation Rate of the CFO Process}
\label{sec:analysis} 

\subsection{Differential Entropy Rates} 
The differential entropy rate of model M2b is: $h_{M2b}=\frac{1}{2}\log_2(2\pi e \omega_c^2 q_2^2 T)$ bits/realization, thanks to $\{n_{AB}\}_k$ being a stationary process with i.i.d. elements, see Eq. (\ref{eq:inc}). The differential entropy rate of model M3 is: $h_{M3}=\log_2(4\pi \Delta)$ bits/realization. Note that $h_{M3}$ is non-negative when $\Delta \ge \frac{1}{4\pi}=0.0796$ Hz which is satisfied easily by the low to medium-end temperature/voltage-controlled oscillators (which culminate in a CFO on the order of hundreds of Hz when tuned to a center frequency of few hundreds of MHz). Also, $h_{M2b}$ is non-negative when $\omega_c^2 T \ge \frac{1}{2\pi e q_2^2} = 1.06\times 10^{16}$ (this inequality is satisfied, say, with $T=1$ ms, for $f_c=\frac{\omega_c}{2\pi}\ge \frac{1.03}{2\pi}$ GHz).

\subsection{Key Generation Rate} 
The key generation rate (KGR) of the proposed method depends on the Auto-correlation function (ACF) of the CFO process $\{f_{AB}\}_k$. Specifically, with $1 \le p \le q$, let $f_{AB}(p)$ and $f_{AB}(q)$ represent the CFO at time $p$ and $q$, respectively. Then, the ACF for model M2a (a first-order ARMA process) is: $ \text{ACF}_{M2a} (p,q)= \sqrt{ \frac{p}{q} } $. It is also straightforward to see that the ACF for model M2b (a stationary process with i.i.d. elements) is: $\text{ACF}_{M2b} (p,q)= \delta(p-q)$ where $\delta(p-q)$ is the Dirac delta function; $\delta(p-q)$ is $1$ for $p=q$, and zero otherwise. Similarly, the ACF for model M3 is: $\text{ACF}_{M3} (p,q)= \delta(p-q)$. For model M2a, a new realization of the CFO process occurs when $\text{ACF}_{M2a} \le \eta$ where $\eta>0$ is a small threshold. Let $T_{M2a}$ denote the time to obtain a new realization for model M2a. Then, $\text{KGR}_{M2a}\le \frac{h_{M2a}}{T_{M2a}}$ bits/sec. For model M2b and model M3, a new realization of the CFO process occurs every $T$ seconds. Therefore, or, $\text{KGR}_{M2b} \le \frac{h_{M2b}}{T}$ bits/sec, and $\text{KGR}_{M3}\le \frac{h_{M3}}{T}$ bits/sec.

\section{Numerical Results}
\label{sec:results}

In this section, we assess the performance of the proposed CFO based secret key generation method by investigating the following metrics: auto-correlation function, key generation rate, key disagreement rate, and decipher probability of Eve.

\begin{figure}[ht]
\begin{center}
	\includegraphics[width=3.5in]{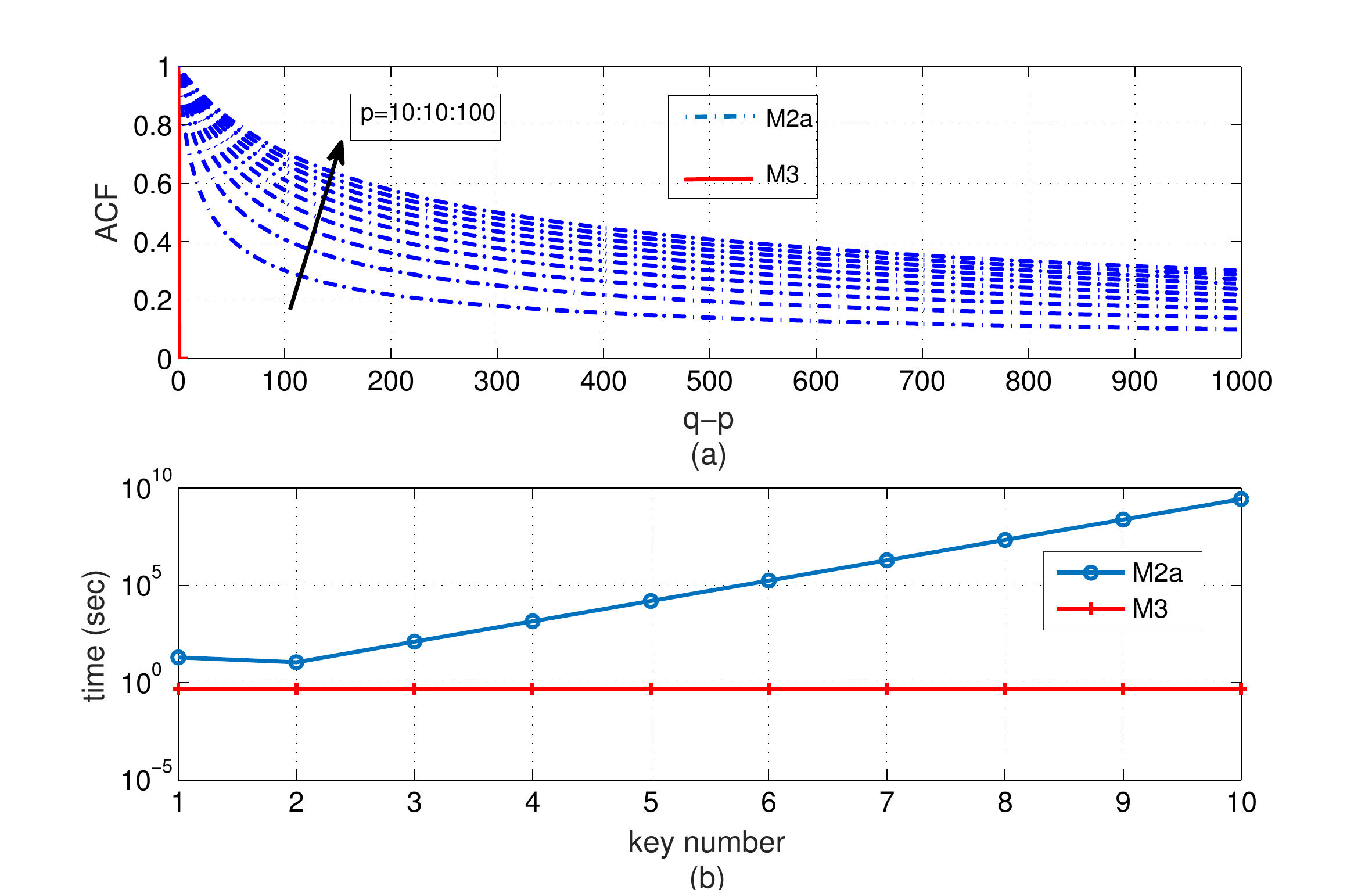} 
\caption{ (a) ACF of the CFO for models M2a, M3, (b) time between generation of two successive keys for models M2a, M3.}
\label{fig:acf}
\end{center}
\end{figure}

Fig. \ref{fig:acf} (a) plots the ACF for the models M2a, and M3. Note that the ACF for model M2a depends explicitly on the absolute time instants $p$ and $q$. Thus, the ACF of model M2a does not decay unless $p$ and $q$ are quite far apart. Therefore, assuming that $f_{AB}(p)$ corresponds to $M$-th sample for $i$-th secret key, and $f_{AB}(q)$ corresponds to first sample for $(i+1)$-th secret key, one can see that the KGR for model M2a decays exponentially over time. For illustration, assume that $T=50$ ms and required $ACF_{M2}< \eta =0.3$. Then, Fig. \ref{fig:acf} (b) shows that the time between generation of two successive keys increases exponentially for model M2a. In other words, due to non-stationary nature of model M2a, the CFO could provide only few ($\sim 10$) secret keys within useful operating time. On the other hand, model M3 could provide 1 secret key every $T$ seconds. In other words, KGR of model M3 is $\frac{1}{T}$ keys/sec.

\begin{figure}[ht]
\begin{center}
	\includegraphics[width=3.5in]{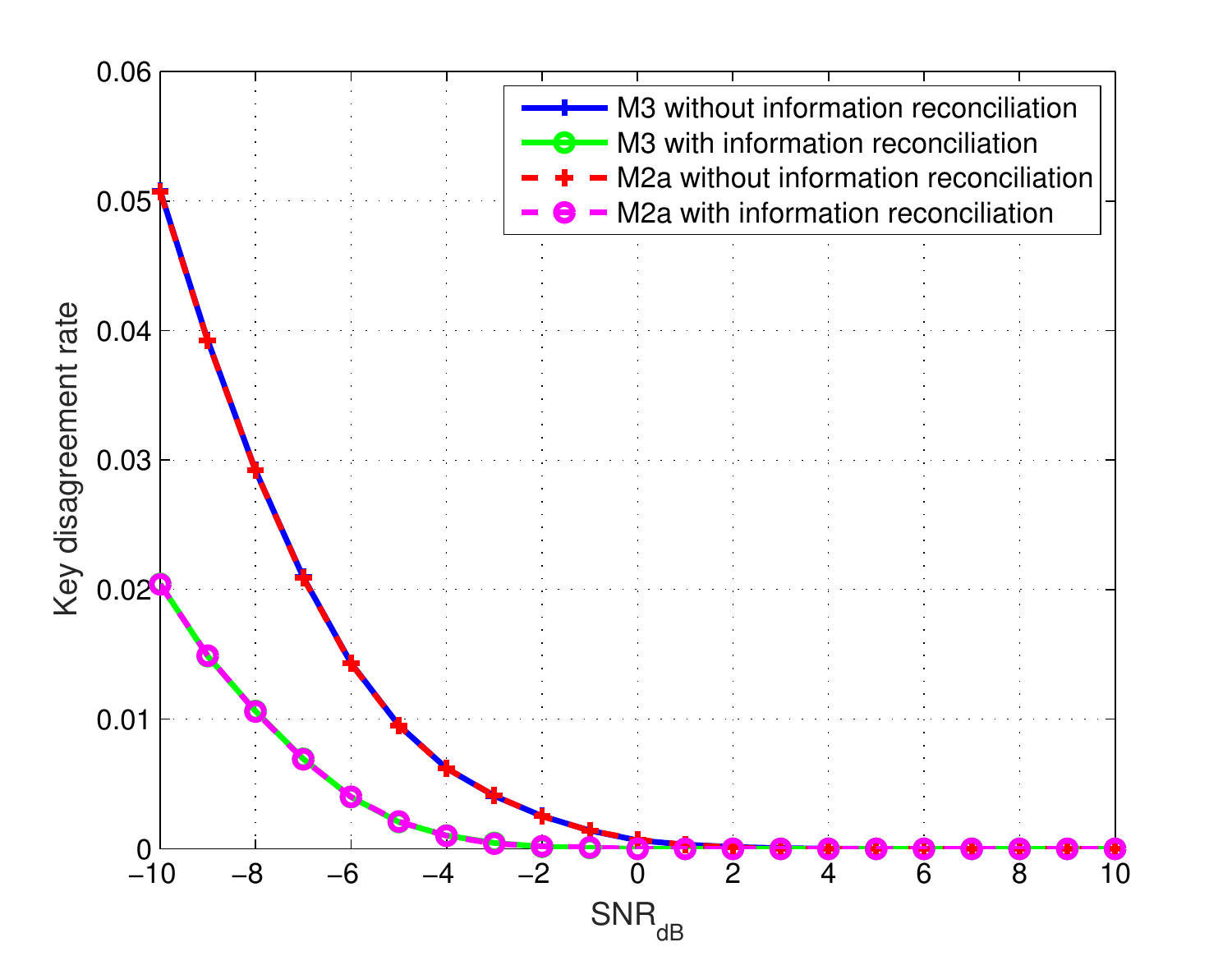} 
\caption{Key disagreement rate vs. SNR.}
\label{fig:kdr}
\end{center}
\end{figure}

Fig. \ref{fig:kdr} plots the average key disagreement rate (KDR)--a measure of the CFO reciprocity--against the signal-to-noise ratio (SNR) for models M2a, and M3. The average KDR is defined as: average number of bits mismatched between the (length-$n$) keys of Alice and Bob. That is, average KDR is computed as $\sum_{N} \frac{\#({K}_A \neq {K}_B)}{n}$, or, $\sum_{N} \frac{\#(\mathcal{K}_A \neq \mathcal{K}_B)}{n}$, depending upon the stage, i.e., before or after information reconciliation. The $\#(A\neq B)$ operator outputs the number of bits mismatched between two length-$n$ sequences $A$ and $B$. Thus, for Monte-Carlo simulations, we set $N=1e5$, use equi-probable quantization with $3$ quantization levels, and utilize Hamming $(7,4)$ code for information reconciliation. Fig. \ref{fig:kdr} reveals that the average KDR decreases exponentially fast with increase in SNR, for both models M2a, M3. Additionally, for any given SNR, the information reconciliation helps reduce the KDR (though the gap diminishes with increase in SNR), as expected. 

\begin{figure}[ht]
\begin{center}
	\includegraphics[width=4in]{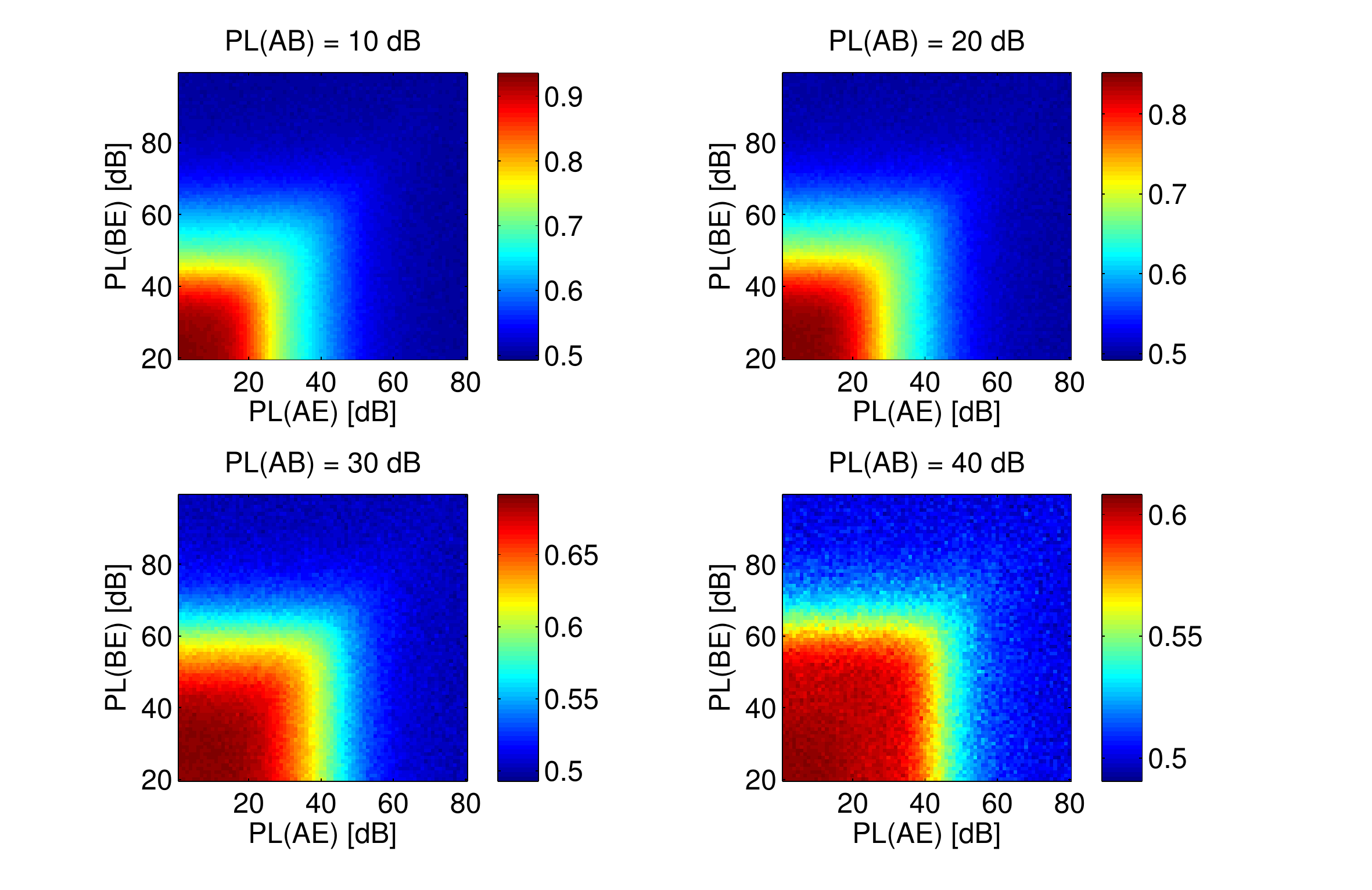} 
\caption{Decipher probability of Eve for a range of pathloss values experienced by the two links seen by the Eve.}
\label{fig:Eve_decipher}
\end{center}
\end{figure}

Fig. \ref{fig:Eve_decipher} plots the average decipher probability of Eve (DPE) as a heat map for a range of pathloss values experienced by the two links (Alice to Eve, and, Bob to Eve) seen by Eve. The average DPE is defined as: average number of bits matched between the (length-$n$) key of Eve and the reconciled keys of Alice and Bob. That is, average DPE is computed as: $\frac{1}{2}\big[\sum_{N} \frac{\#({K}_E = \mathcal{K}_A)}{n}+\sum_{N} \frac{\#({K}_E = \mathcal{K}_B)}{n}\big]$ where $K_E$ is the key at Eve. $K_E$ was constructed by invoking the step 2 of the proposed method on $\hat{\omega}_{AE}-\hat{\omega}_{BE}$ (the Eve's belief about the shared secret key). One could see that the average DPE decreases from 0.9 to 0.5 as soon as either of the two links observed by the Eve becomes more degraded compared to the link between Alice and Bob.

\section{Conclusion}
\label{sec:conclusion}

We have proposed to utilize the reciprocal CFO to generate shared secret keys between a legitimate node pair in the presence of a malicious node. Simulation result have shown that the KDR decreases exponentially with increase in the SNR of the link between Alice and Bob. Furthermore, the average DPE decreases as soon as either of the two links observed by the Eve becomes more degraded compared to the link between Alice and Bob. We have also computed the KGR of the CFO process for the two frequency-stability regimes of oscillators. 

Some comments about the proposed method are in order. CFO based key generation is appealing because CFO estimation is easily carried out, and already a mandatory operation for the modern cellular/WiFi receivers. Also, the average DPE could approach to zero when the legitimate node pair employs multiple-antenna/beamforming techniques to ensure that minimum power is radiated in unintended directions. Finally, high frequency bands such as milli-meter wave/$60$ GHz band and terahertz band could benefit from the proposed method because the KGR of the proposed method is proportional to the center frequency of operation. In short, the proposed method could act as first line of defense against the malicious nodes who are either facing degraded/bad channels, or, don't have the computational resources for sophisticated, real-time signal processing. In near future, we aim to prototype the proposed algorithm on GNU radio/USRP based SDR platform.

\appendices

\section{Blind CFO estimation from BPSK waveform}
\label{app:blind_est}

The BPSK baseband waveform at transmitter is: $x(t) = \sum_k a_k p(t-kT)$ where $a_k \in \{ 1,-1\}$, $p(t)$ is the pulse shape and $T$ is symbol duration. Then, the signal received at the receiver is: $y(t) = x(t)\exp(j2\pi \Delta f t)$. To estimate the CFO $\Delta f$, one needs to perform a series of operations on $y(t)$. Specifically,

\begin{eqnarray*}
y^{2}(t) & = & \{x(t)\exp(j2\pi \Delta f t)\}^2 \\
& = & x^2(t) \exp(j4\pi \Delta f t) \\
& = & \{\sum_k a_k p(t-kT)\}^2 \exp(j4\pi \Delta f t) \\
& = & \{\sum_k p^2(t-kT)\} \exp(j4\pi \Delta f t) \\
\end{eqnarray*}
Let $f(t)=\sum_k p^2(t-kT)$. Then one can write:
\begin{equation}
f(t) = \frac{a_0}{2}+ \sum_{n=1}^{\infty} ( a_n \cos \frac{2n\pi t}{T} + b_n \sin \frac{2n\pi t}{T} )
\end{equation}
where $a_0$, $a_n$ and $b_n$ are the Fourier series coefficients given as: $a_0 = \frac{2}{T} \int_0^T f(t) \, dt $, $a_n = \frac{2}{T} \int_0^T f(t) \cos(\frac{2n\pi t}{T})\, dt $, and $b_n = \frac{2}{T} \int_0^T f(t) \sin(\frac{2n\pi t}{T})\, dt $. Then,
\begin{eqnarray*}
y^{2}(t) & = & f(t) \exp(j4\pi \Delta f t) \\
& = &  \{ \frac{a_0}{2}+ \sum_{n=1}^{\infty} ( a_n \cos \frac{2n\pi t}{T} + b_n \sin \frac{2n\pi t}{T} ) \} \exp(j4\pi \Delta f t) \\
\end{eqnarray*}
Let $\frac{1}{T}=F_{sym}$, the symbol rate. Let us now write the real and imaginary parts of $y^2(t)$ separately:
\begin{equation} \label{realysq}
\begin{split}
\Re (y^2(t)) &= \frac{a_0}{2}\cos(4\pi \Delta ft) \\
           &+\sum_{n=1}^{\infty} ( a_n \cos (2\pi nF_{sym}t)\cos(4\pi \Delta ft) \\
           &+ b_n \sin (2\pi nF_{sym}t)\cos(4\pi \Delta ft) )
\end{split}
\end{equation}
\begin{equation} \label{imagysq}
\begin{split}
\Im (y^2(t)) &= \frac{a_0}{2}\sin(4\pi \Delta ft) \\
           &+\sum_{n=1}^{\infty} ( a_n \cos (2\pi nF_{sym}t)\sin(4\pi \Delta ft) \\
           &+b_n \sin (2\pi nF_{sym}t)\sin(4\pi \Delta ft) )
\end{split}
\end{equation}
 
Passing the complex-valued signal $y^2(t)$ of Eqs. (\ref{realysq}), (\ref{imagysq}) through a low-pass filter with cut-off frequency $\omega_c$ in the range $4\pi \Delta f$ $<<$ $\omega_c$ $<<$ $2\pi F_{sym}$, we get:
\begin{equation}
z(t) = \frac{a_0}{2} \exp(j4\pi \Delta ft)
\end{equation}  
Finally, we take the fast Fourier transform (FFT) of $z(t)$, find the frequency $\widehat{2\Delta f}$ corresponding to the peak value, and divide it by 2 which gives us the CFO estimate.


\footnotesize{
\bibliographystyle{IEEEtran}
\bibliography{references}

\begin{thebibliography}{10}
\providecommand{\url}[1]{#1}
\csname url@rmstyle\endcsname
\providecommand{\newblock}{\relax}
\providecommand{\bibinfo}[2]{#2}
\providecommand\BIBentrySTDinterwordspacing{\spaceskip=0pt\relax}
\providecommand\BIBentryALTinterwordstretchfactor{4}
\providecommand\BIBentryALTinterwordspacing{\spaceskip=\fontdimen2\font plus
\BIBentryALTinterwordstretchfactor\fontdimen3\font minus
  \fontdimen4\font\relax}
\providecommand\BIBforeignlanguage[2]{{%
\expandafter\ifx\csname l@#1\endcsname\relax
\typeout{** WARNING: IEEEtran.bst: No hyphenation pattern has been}%
\typeout{** loaded for the language `#1'. Using the pattern for}%
\typeout{** the default language instead.}%
\else
\language=\csname l@#1\endcsname
\fi
#2}}

\bibitem{shannon:BSTJ:1949}
C.~E. Shannon, ``Communication theory of secrecy systems,'' \emph{Bell system
  technical journal}, vol.~28, no.~4, pp. 656--715, 1949.

\bibitem{Wyner:BLTJ:1975}
\BIBentryALTinterwordspacing
A.~D. Wyner, ``The wire-tap channel,'' \emph{Bell System Technical Journal},
  vol.~54, no.~8, pp. 1355--1387, 1975. [Online]. Available:
  \url{http://dx.doi.org/10.1002/j.1538-7305.1975.tb02040.x}
\BIBentrySTDinterwordspacing

\bibitem{Csiszar:TIT:1978}
I.~Csiszar and J.~Korner, ``Broadcast channels with confidential messages,''
  \emph{Information Theory, IEEE Transactions on}, vol.~24, no.~3, pp.
  339--348, 1978.

\bibitem{Maurer:TIT:1993}
U.~Maurer, ``Secret key agreement by public discussion from common
  information,'' \emph{Information Theory, IEEE Transactions on}, vol.~39,
  no.~3, pp. 733--742, 1993.

\bibitem{Amitav:corr:2010}
A.~Mukherjee, S.~A.~A. Fakoorian, J.~Huang, and A.~L. Swindlehurst,
  ``Principles of physical layer security in multiuser wireless networks: A
  survey,'' \emph{CoRR}, vol. abs/1011.3754, 2010.

\bibitem{Shiu:WC:2011}
Y.-S. Shiu, S.-Y. Chang, H.-C. Wu, S.-H. Huang, and H.-H. Chen, ``Physical
  layer security in wireless networks: a tutorial,'' \emph{Wireless
  Communications, IEEE}, vol.~18, no.~2, pp. 66--74, 2011.

\bibitem{Bloch:TIT:2008}
M.~Bloch, J.~Barros, M.~R.~D. Rodrigues, and S.~McLaughlin, ``Wireless
  information-theoretic security,'' \emph{Information Theory, IEEE Transactions
  on}, vol.~54, no.~6, pp. 2515--2534, 2008.

\bibitem{Bloch:ITW:2006}
M.~Bloch, A.~Thangaraj, S.~McLaughlin, and J.~M. Merolla, ``Ldpc-based gaussian
  key reconciliation,'' in \emph{Information Theory Workshop, 2006. ITW '06
  Punta del Este. IEEE}, 2006, pp. 116--120.

\bibitem{Bennett:TIT:1995}
C.~Bennett, G.~Brassard, C.~Crepeau, and U.~Maurer, ``Generalized privacy
  amplification,'' \emph{Information Theory, IEEE Transactions on}, vol.~41,
  no.~6, pp. 1915--1923, 1995.

\bibitem{Ye:TIFS:2010}
C.~Ye, S.~Mathur, A.~Reznik, Y.~Shah, W.~Trappe, and N.~B. Mandayam,
  ``Information-theoretically secret key generation for fading wireless
  channels,'' \emph{Information Forensics and Security, IEEE Transactions on},
  vol.~5, no.~2, pp. 240--254, 2010.

\bibitem{Ye:ISIT:2006}
C.~Ye, A.~Reznik, and Y.~Shah, ``Extracting secrecy from jointly gaussian
  random variables,'' in \emph{Information Theory, 2006 IEEE International
  Symposium on}, 2006, pp. 2593--2597.

\bibitem{Dong:TSP:2010}
L.~Dong, Z.~Han, A.~Petropulu, and H.~Poor, ``Improving wireless physical layer
  security via cooperating relays,'' \emph{Signal Processing, IEEE Transactions
  on}, vol.~58, no.~3, pp. 1875--1888, 2010.

\bibitem{Golla:infocom:2011}
S.~Gollakota and D.~Katabi, ``Physical layer wireless security made fast and
  channel independent,'' in \emph{INFOCOM, 2011 Proceedings IEEE}, 2011, pp.
  1125--1133.

\bibitem{Lim:VLSI:2005}
D.~Lim, J.~Lee, B.~Gassend, G.~Suh, M.~van Dijk, and S.~Devadas, ``Extracting
  secret keys from integrated circuits,'' \emph{Very Large Scale Integration
  (VLSI) Systems, IEEE Transactions on}, vol.~13, no.~10, pp. 1200--1205, 2005.

\bibitem{Wang:ICC:2012}
W.~Hou, X.~Wang, and J.~Chouinard, ``Physical layer authentication in ofdm
  systems based on hypothesis testing of cfo estimates,'' in
  \emph{Communications (ICC), 2012 IEEE International Conference on}, 2012, pp.
  3559--3563.

\bibitem{Mahboob:Globecom:2014}
M.~M.~U. Rahman, A.~Yasmeen, and J.~Gross, ``Phy layer authentication via
  drifting oscillators,'' in \emph{2014 IEEE Global Communications Conference},
  Dec 2014, pp. 716--721.

\bibitem{Mahboob:ICUWB:2015}
M.~M.~U. Rahman, S.~Kanwal, and J.~Gross, ``Simultaneous energy harvesting and
  sender-node authentication at a receiver node,'' in \emph{2015 IEEE
  International Conference on Ubiquitous Wireless Broadband (ICUWB)}, Oct 2015,
  pp. 1--5.

\bibitem{Satyam:Arxiv:2017}
S.~Dwivedi, J.~O. Nilsson, P.~Papadimitratos, and P.~H{\"a}ndel, ``Climex: A
  wireless physical layer security protocol based on clocked impulse
  exchanges,'' \emph{arXiv preprint arXiv:1708.04774}, 2017.

\bibitem{Imai:JWIS:2006}
H.~Imai, K.~Kobara, and K.~Morozov, ``On the possibility of key agreement using
  variable directional antenna,'' 2006.

\bibitem{Mahboob:VTC:2017S}
M.~M.~U. Rahman, A.~Yasmeen, and Q.~H. Abbasi, ``Exploiting lack of hardware
  reciprocity for sender-node authentication at the phy layer,'' in \emph{2017
  IEEE 85th Vehicular Technology Conference (VTC Spring)}, June 2017, pp. 1--5.

\bibitem{Zucca:ITUFFC:2005}
C.~Zucca and P.~Tavella, ``The clock model and its relationship with the allan
  and related variances,'' \emph{Ultrasonics, Ferroelectrics and Frequency
  Control, IEEE Transactions on}, vol.~52, no.~2, pp. 289--296, 2005.

\bibitem{Galleani:Metrologia:2008}
L.~Galleani, ``A tutorial on the two-state model of the atomic clock noise,''
  \emph{Metrologia}, vol.~45, no.~6, p. S175, 2008.

\bibitem{Quitin:TWC:2013}
F.~Quitin, M.~M.~U. Rahman, R.~Mudumbai, and U.~Madhow, ``A scalable
  architecture for distributed transmit beamforming with commodity radios:
  Design and proof of concept,'' \emph{Wireless Communications, IEEE
  Transactions on}, vol.~12, no.~3, pp. 1418--1428, 2013.

\end{thebibliography}
}

\vfill\break

\end{document}